\documentclass{PoS}

\usepackage{multirow, graphicx, epsfig, color}
\DeclareGraphicsExtensions{.eps}

\def\teq#1{$\, #1\,$}                         

\font\fiverm=cmr5

          \font\sixrm=cmr6

\def\gammax{\gamma_{\rm max}}

\def\ecyc{\omega_{\hbox{\fiverm B}}}
\def\wp{\omega_{\hbox{\sixrm p}}}

\def\acknowledgements#1{\noindent{\bf Acknowledgements}: #1}

\title{Time-Dependent, Multi-Wavelength Shock Acceleration Models for Active Flares of 3C 279}

\ShortTitle{Time-Dependent Shock Models for Active Flares of 3C 279}

\author{\speaker{Matthew G. Baring}\\ 
        Department of Physics and Astronomy - MS 108, Rice University,
        6100 Main Street, Houston, Texas 77251-1892, USA\\
        E-mail: \email{baring@rice.edu}}

\author{Markus B\"{o}ttcher\\
        Centre for Space Research, North-West University, Potchefstroom, 2531, South Africa\\
        E-mail: \email{Markus.Bottcher@nwu.ac.za}}

\abstract{
Jets in blazars are an excellent forum for studying acceleration at
relativistic shocks using the highly-variable emission seen across the
electromagnetic spectrum.  Our recent work on combining multi-wavelength
leptonic emission models with simulated thermal+non-thermal distributions from
shock acceleration theory has resulted in new insights into plasma
conditions in blazars. This has demonstrated the ability to infer the
cyclotron frequency, the plasma density and thus also the Alfven speed,
thereby determining the rapidity of particle energization.  An important
inference was that MHD turbulence levels decline with remoteness from
jet shocks.  This paper outlines new results from our recent
extension of this program to a two-zone, time-evolving construction,
modeling together both extended, enhanced emission states from larger
radiative regions, and prompt flare events from compact acceleration
zones. These are applied to flares in the FSRQ blazar 3C 279 monitored by
{\it Fermi}-LAT in gamma-rays in late 2013. 
With impulsive injection episodes from the shock zone, as the
acceleration first proceeds and then abates, the radiative simulations
obtain a pronounced spectral hardening in the optical and gamma-ray bands
as the flare grows, followed by a softening during the decay phase.  
For 3C 279, while model radio and X-ray
synchrotron flares are temporally correlated, there is a lag in both bands
relative to GeV gamma rays and optical emission on timescales of a number of
hours.   This delay is governed by the short cooling time associated with the
bright external Compton signal.  
}

\FullConference{High Energy Phenomena in Relativistic Outflows VII - HEPRO VII\\
		9-12 July 2019\\
		Facultat de F\'{i}sica, Universitat de Barcelona, Spain}

\begin{document}

\section{Introduction}

Extragalactic jet sources associated with active galactic nuclei (AGN)
are among the most powerful sites for radiation 
in the cosmos.  The jets are highly-collimated outflows, imaged in radio, optical and 
X-ray in select examples, and can emit up to around \teq{10^{47}}erg/sec in 
highly variable radiation, spanning wavelength ranges from the radio to
TeV gamma-rays.  The sub-class of blazars, the most variable of AGN, 
are believed to have their jets aligned near to the line of sight to Earth.
Current understanding of blazar jets is limited by the fact that the
physical phenomena of jet formation and collimation, particle
acceleration, and radiation, are usually treated as separate problems.
The acceleration/radiation interface, the emphasis of this paper, can be suitably addressed using
plasma (particle-in-cell; PIC \cite{Nishikawa-2005-ApJ,Sironi-2009-ApJ}) 
and diffusion (Monte Carlo) codes \cite{Ellison-1990-ApJ,SB12}.

To help bridge the gap between science focal areas and theoretical techniques, 
we embarked upon a program to blend shock acceleration and
radiation elements of blazar models \cite{BBS17}. The acceleration process was modeled 
using the relativistic Monte Carlo simulation shock study of \cite{SB12}, with the 
principal product being combined thermal plus non-thermal distributions spanning 
\teq{3-7} decades in particle momentum.  These constitute leptons injected 
in a fairly confined acceleration zone in the 
jet, which are then subject to radiative dissipation in larger, neighboring jet regions.  
The radiation modules of \cite{Boettcher-2002-ApJ} and \cite{Boettcher-2013-ApJ}
were then employed to characterize the multi-wavelength (MW) synchrotron and inverse 
Compton signatures. Such broadband modeling affords tighter constraints on jet environmental 
parameters, including the magnetic field strength (i.e., \teq{\ecyc = eB/m_ec}), 
the electron density \teq{n_e} and its plasma frequency \teq{\wp = [4\pi n_e e^2/m_e]^{1/2}}, 
and the nature and spatial distribution of the field turbulence.   To efficiently inject
electrons from thermal energies into the acceleration process, \cite{BBS17} determined that the 
shock needed fairly turbulent fields on small scales, such as would be generated by the 
Weibel instability: see \cite{Nishikawa-2005-ApJ,Sironi-2009-ApJ} for PIC simulations.  
To position the synchrotron turnover in the optical, 
very large mean free paths and inefficient diffusion/acceleration were needed for the 
highest energy electrons (first concluded by \cite{Inoue-1996-ApJ}), 
{\it corresponding to very low levels of turbulence}.  
Deducing such disparate character in field turbulence 
spanning the large range in momenta of accelerated particles signals the power of 
this synergistic approach.

Since the introductory steady-state foray of \cite{BBS17}, we have moved our program forward 
by exploring time-dependent properties of energetic electron populations in combination 
with MW radiation signals, encapsulating the essentials of the competition between shock acceleration 
and radiative cooling.  This has culminated in an extensive offering \cite{BB19_ApJ} 
on evolving MW SEDs and light curves for blazar flares, 
of which this paper captures a glimpse pertaining to the famous flat spectrum radio quasar (FSRQ)
3C 279, a low synchrotron-energy-peaked (LSP) blazar.

\section{Shock Acceleration and Multi-Wavelength Radiation Emission}

Electron distributions for the generation of radiation signatures are obtained in 
our study using the shock acceleration Monte Carlo simulation of \cite{SB12}.  
This code tracks the diffusive elements of the first-order Fermi acceleration process (DSA), 
modeling stochastic pitch angle diffusion of charges convecting along magnetic field lines,
yet it also includes episodes of shock drift energization (SDA).  
The mean-free path for pitch angle scattering is parameterized via
\teq{\lambda_{\rm pas}= \eta (p) \, r_{\rm g}}, i.e. a momentum-dependent multiple 
\teq{\eta (p)} of the particle's gyro radius, \teq{r_{\rm g} = p c / (q B)}, where $p$ is the particle's momentum. 
A broadly applicable choice for the scaling is a power-law in the particle's momentum, \teq{\eta (p) = \eta_1 \, (p/mc)^{\alpha - 1}}, 
where \teq{\eta_1} describes the mean free path in the non-relativistic 
limit, \teq{\gamma \to 1}.  Motivations for this form from quasi-linear MHD turbulence theory,
hybrid plasma simulations, and {\it in-situ} spacecraft observations in the heliosphere are discussed in \cite{BBS17,SB12}.
Examples of simulated distributions for strong, subluminal, 
mildly-relativistic shocks expected in blazar jets are provided in Fig.~1 of \cite{BBS17}, illustrating that shock acceleration 
leads to a non-thermal broken power-law tail of relativistic charges 
that have been accelerated out of the thermal pool.  
As a consequence of the \teq{\eta (p) \propto p^{\alpha -1}} form, 
the particle distribution is somewhat steep 
(\teq{dn/dp \sim p^{-2.2}}) at low momenta when DSA dominates, and much flatter
(\teq{dn/dp \sim p^{-1}}) for much higher momenta when SDA is the more effective energization process.

These thermal + non-thermal distributions, modulated by an exponential radiation-reaction-limited turnover 
at a maximum Lorentz factor \teq{\gammax} (see \cite{BB19_ApJ}), serve as an injection 
\teq{Q_e(\gamma ,\, t)} into a time-dependent electron/pair evolution code.  The electrons 
then diffuse in a larger radiation zone, surrounding the shock region
(see Fig.~2 of \cite{BBS17}), cooling while radiating.  This evolving distribution of relativistic electrons 
is simulated by numerically solving a Fokker-Planck equation of the form
\begin{equation}
    {\partial n_e (\gamma_e, t) \over \partial t} = - {\partial \over \partial \gamma_e} \Bigl( \dot\gamma_e
    \, n_e [\gamma_e, t] \Bigr) - {n_e (\gamma_e, t) \over t_{\rm esc, e}} + Q_e (\gamma_e, t) 
  \label{eq:FP}
\end{equation}
using an implicit Crank-Nicholson scheme \cite{Boettcher-2002-ApJ}. 
Here $\dot\gamma_e$ represents the combined radiative energy loss rate of the electrons.
The relevant radiative mechanisms are synchrotron emission in a tangled magnetic field,
synchrotron self-Compton (SSC) radiation, and inverse Compton scattering of external radiation 
fields (external inverse Compton = EIC) on various plausible target photon fields such as from 
a dusty torus proximate to the jet.  The electron distribution \teq{n_e (\gamma_e, t)} is 
considered in the co-moving frame of the jet, and is assumed isotropic therein, with an 
electron escape time scale parameterized via
a multiple of the light-crossing time scale, \teq{t_{\rm esc, e} = \eta_{\rm esc} \, R/c}.  

The spatial transfer and temporal evolution of the jet frame photon distribution 
\teq{n_{\rm ph} (\epsilon, t)} is modeled by solving a continuity equation
\begin{equation}
    {\partial n_{\rm ph} (\epsilon, t) \over \partial t} = {4 \, \pi \, j_{\epsilon} \over \epsilon \, m_e c^2} 
    - c \, \kappa_{\epsilon} \, n_{\rm ph} (\epsilon, t) - {n_{\rm ph} (\epsilon, t) \over t_{\rm esc, ph}}
  \label{eq:radiation}
\end{equation}
that serves as an upgrade from the steady-state radiative transfer routines of \cite{Boettcher-2013-ApJ} 
that were employed in \cite{BBS17}.
Here \teq{j_{\epsilon}} and \teq{\kappa_{\epsilon}} are the emissivity and absorption coefficients, respectively,
\teq{\epsilon = h \nu / (m_e c^2)} is the dimensionless photon energy, and \teq{t_{\rm esc, ph}} is the photon escape
time scale from the radiation emission region. 
Note that \teq{\gamma\gamma} pair absorption is generally small in the jet for 3C 279 due to its 
low IC peak energy in the GeV band.  Note also that we invoke a dusty torus seed for 
the external inverse Compton (EIC) component that is Doppler-boosted and highly anisotropic in the jet frame, thereby 
strongly enhancing the EIC emissivity. Corrections for \teq{\gamma\gamma} absorption by the Extragalactic Background
Light are included but are small.  The total observed flux is obtained by Doppler boosting \teq{n_{\rm ph} (\epsilon, t)} 
to the observer frame, with 
jet-frame and observer time intervals being related through \teq{\Delta t_{\rm obs} = \Delta t \, (1 + z) / \delta}, 
for a redshift of $z = 0.536$. Our radiation emission/transfer code outputs snap-shot SEDs and multi-wavelength light 
curves at select frequencies.

\section{Modeling Two Contrasting Flares from 3C 279 in December 2013}

Our focal blazar, 3C~279, is  one of the brightest gamma-ray blazars detected by the Large Area Telescope (LAT) on board the {\it Fermi}
Gamma-Ray Space Telescope (e.g., \cite{Abdo10}).  It is one of only a few FSRQs also detected in
very-high-energy (VHE: $E > 100$~GeV) gamma rays by ground-based Imaging Atmospheric \v{C}erenkov Telescopes
(e.g., \cite{Albert-2008-Sci}).  The active period of interest was December 2013 -- April 2014,
during which extensive multi-wavelength observations of flaring activity were acquired, as reported in
\cite{Hayashida15}. Figure 7 of that paper shows multi-wavelength light curves of 3C~279, where several
gamma-ray flares (B, C, D) are identified, in addition to a quiescent period (A).   In this presentation, we
summarize our modeling of two of these ephemerals, the modest Flare C, and the strongly Compton-dominated Flare B, 
an investigation presented at length in \cite{BB19_ApJ} (see also \cite{BB19}).
In the modeling, the jet Doppler factor was presumed to be $\delta \approx 15$, 
and the variability timescales of the order of hours imply that the active regions cannot be larger than $\sim 10^{16}$cm.

\begin{figure}[hb]
\centering
\vspace{-12pt}
\centerline{\includegraphics[width=7.8cm]{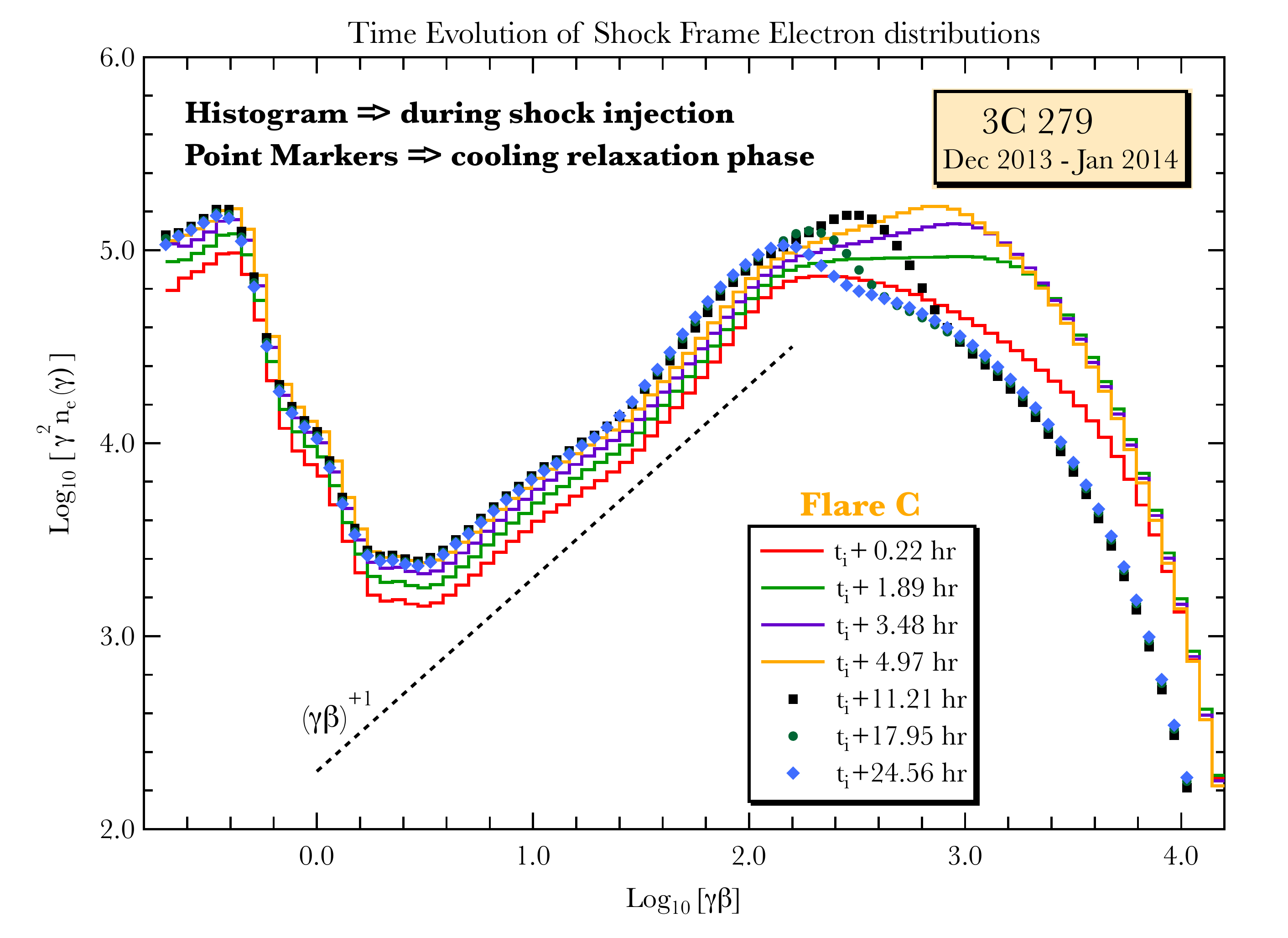}
 \hskip -8pt \includegraphics[width=8.0cm]{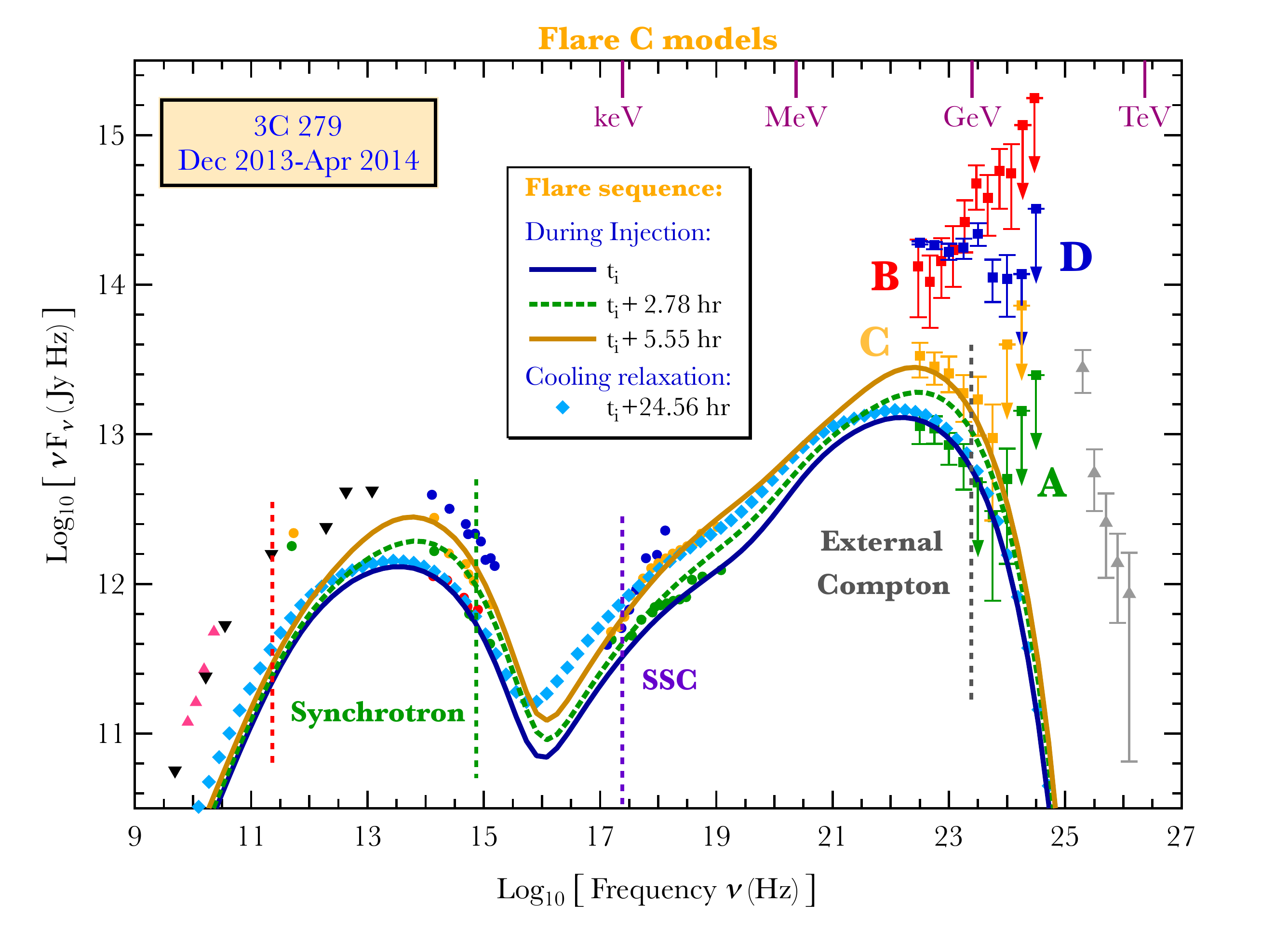}}
\vspace{-15pt}
\caption{{\it Left panel}: 
Simulated electron distribution sequence \teq{n_e(\gamma_e, \, t)}
corresponding to the Flare C spectral sequence on the right. 
The diagonal line marks the approximate shock injection power law distribution that 
results primarily from the shock drift (SDA) mechanism: see \cite{BBS17}.
{\it Right panel}: 
Snap-shot SEDs of 3C279 modeling the moderate Flare C on December 31, 2013
at different times.  Data are from \cite{Hayashida15}. Curves illustrate
the spectral evolution during the rising part of Flare C, 
the injection phase at times \teq{t< t_i + 5.55}hr; the acqua diamond spectrum 
signals the evolution during the decaying part. See text for details. }
 \label{fig:FlareC_edist_spec}
\end{figure}

Fig.~\ref{fig:FlareC_edist_spec} shows the electron distributions (at left) resulting from the 
solutions of the kinetic equation in Eq.~(\ref{eq:FP}), together with the MW radiation modeling (at right)
associated with Eq.~(\ref{eq:radiation}) for Flare C.  The spectra are temporal snap-shots 
at different times, as labeled, and are compared with observations extracted from \cite{Hayashida15}.
The jet/shock parameters for this modeling are given in Tables~2 and~3 of \cite{BB19_ApJ}.  
We highlight here that the shock acceleration diffusion parameters for the Flare C models
were \teq{\eta_1=100} and \teq{\alpha =3}, i.e. that the pitch-angle scattering mean free path 
scaled as \teq{\lambda_{\rm pas} = 100 \, r_g \, (p/m_ec)^2\, \propto p^3}.  
These parameters were realized throughout the flare, 
including at the outset, which is represented by the dark blue curve at right, and constitute
the long-term equilibrium state prior to the flare (period A).  Thus while the injection rate became enhanced during 
the flare, the character of the turbulence in the shock environs was not altered.
In particular, the \teq{\lambda_{\rm pas}/r_g\gg 1} realization for essentially all electron 
momenta implies that MHD turbulence levels are extremely low on the pertinent diffusion scales,
a necessary imposition for the synchrotron emission to emerge in the optical for this LSP blazar:
see \cite{Inoue-1996-ApJ,BBS17} for explanation of this constraint.  The connection between 
turbulence and polarization measures \cite{Hayashida15} is discussed in \cite{BB19_ApJ}.

The electron distributions on the left of Fig.~\ref{fig:FlareC_edist_spec} evince 
flat \teq{n_e(\gamma ) \equiv dn/dp \sim p^{-1}} character at low momenta 
(\teq{\gamma\beta \lesssim 30}; note the \teq{\gamma^2 n_e(\gamma )} representation)
due to the prominent action of SDA.  The densities rise due to the cumulative injection 
during the 5.5 hour activation period.  At high energies, acceleration first competes 
with cooling, generating electrons with \teq{\gamma\gtrsim 10^3}, and then 
succumbs to it at \teq{t>5.5}hrs when the injection is terminated.  Thereafter 
the cooling evolution of \teq{n_e(\gamma)} is apparent, causing a pile-up 
of electrons at \teq{\gamma\sim 10^2}.  The associated radiation signatures 
on the right of Fig.~\ref{fig:FlareC_edist_spec} consist of a simple rise during 
the acceleration epoch, followed by a cooling/spectral softening in 
the optical synchrotron and EIC GeV $\gamma$-ray bands on timescales of 
\teq{\sim 10}hr, consistent with the {\it Fermi}-LAT light curve in \cite{Hayashida15}.
The EIC component seeded by IR radiation from a dusty torus is needed because the low \teq{\gammax} 
required to generate a synchrotron peak in the optical produces an SSC peak in the hard X-rays.
The radio synchrotron and SSC X-ray fluxes do not die off as quickly during the cooling relaxation phase, 
a signature that is borne out in the model light curves for Flare C displayed in 
Fig.~4 of \cite{BB19_ApJ} that were derived for the 4 frequencies marked 
in Fig.~\ref{fig:FlareC_edist_spec}.  
Note that the steady radio emission is believed to originate from a region much larger
than that for the flares, and so does not provide a useful constraint for the MW data ``fitting'' protocol.

\begin{figure}[ht]
\centering
\centerline{\includegraphics[width=8.0cm]{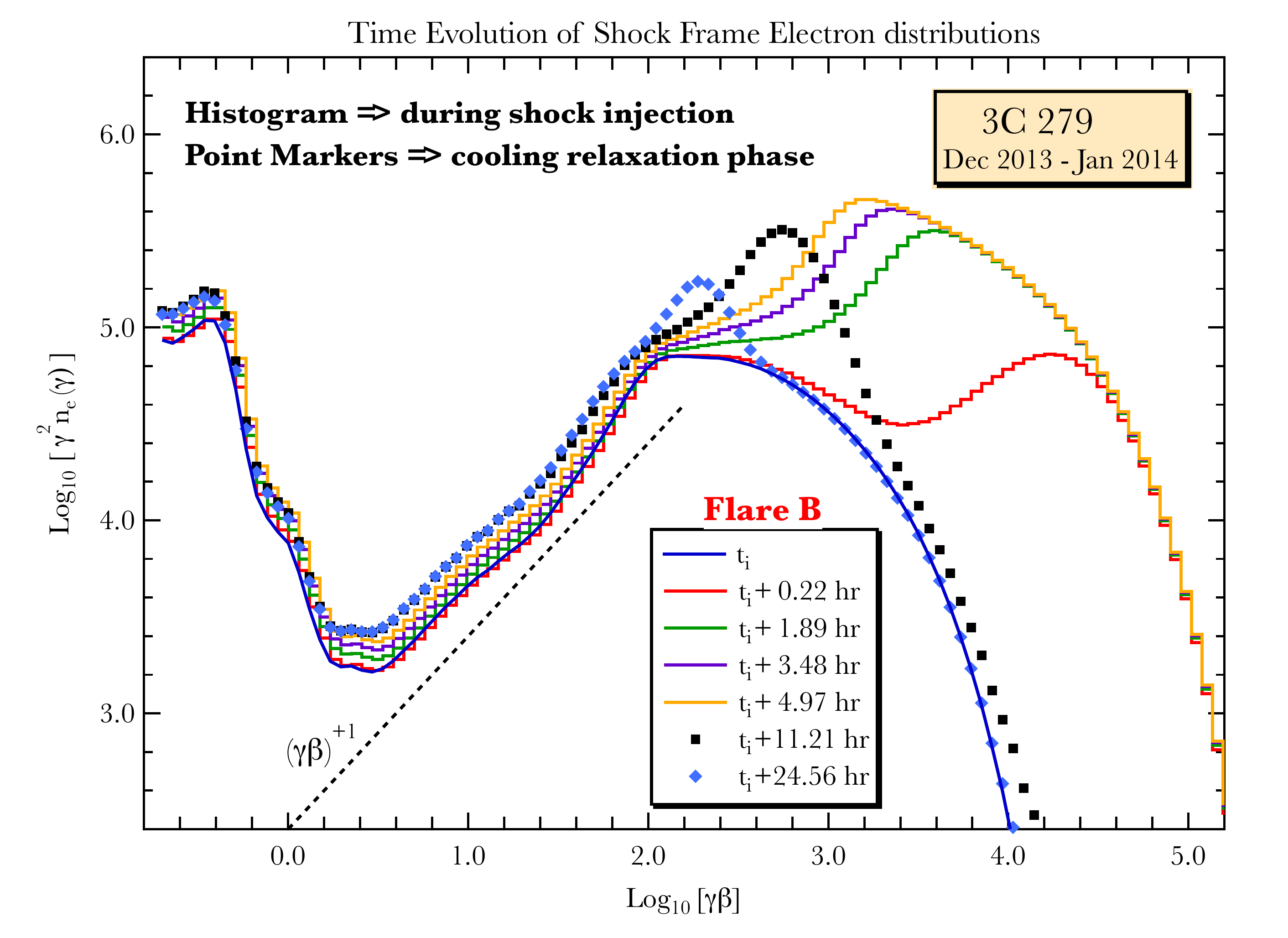}
 \hskip -8pt \includegraphics[width=8.2cm]{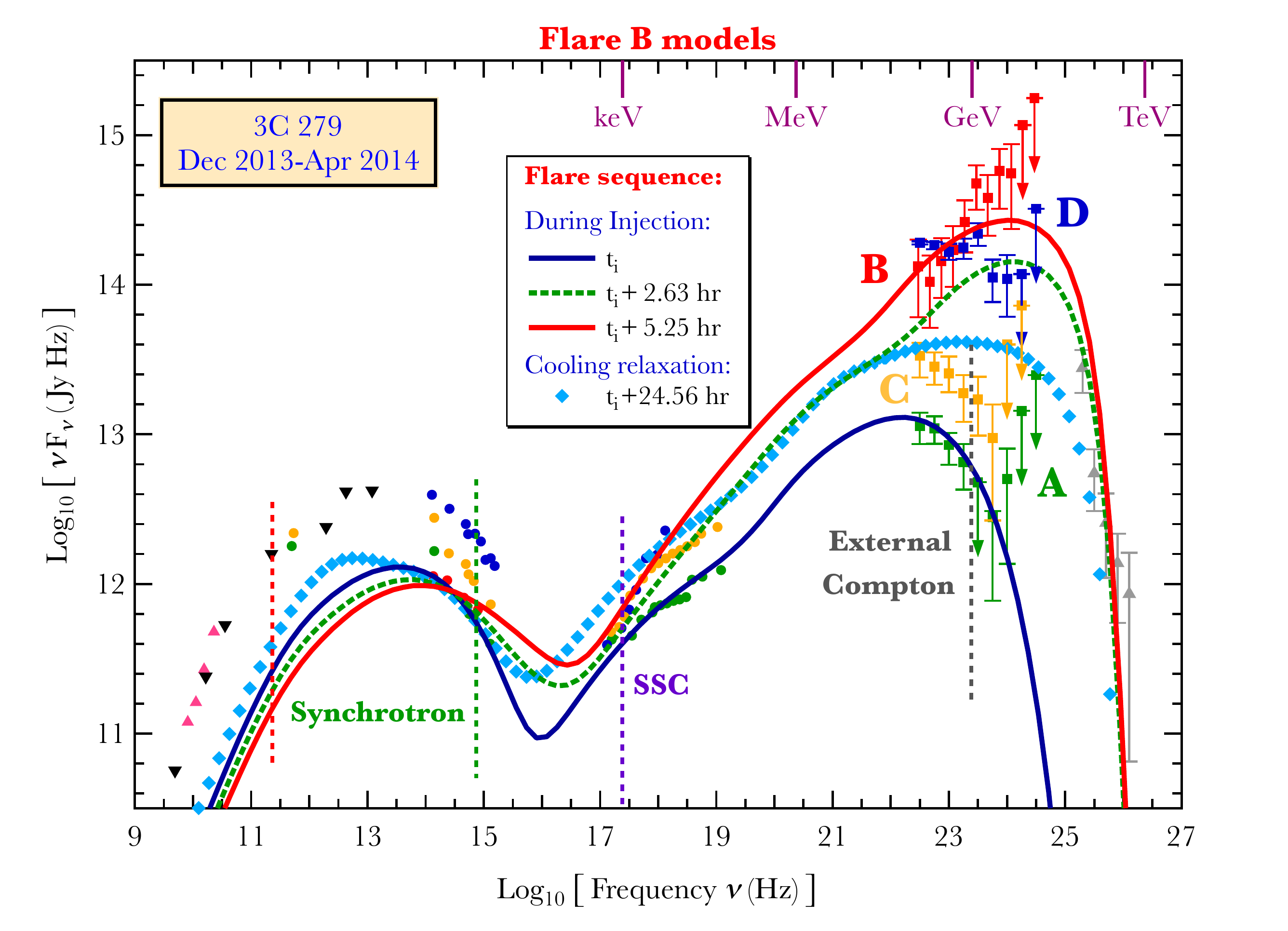}}
\vspace{-15pt}
\caption{{\it Left panel}: 
Simulated electron distribution sequence \teq{n_e(\gamma_e, \, t)}
corresponding to the Flare B spectral sequence on the right. 
The solid blue curve is the long-term equilibrium distribution.  As in Fig.~1, 
the energetic injection from particle acceleration and subsequent cooling are apparent.
{\it Right panel}: 
Snap-shot SEDs of 3C279 modeling the moderate Flare B on December 20, 2013.
Data are again from \cite{Hayashida15}. As in Fig.~1, 
the curves illustrate the spectral evolution during the rising part of Flare B, 
the injection phase at times \teq{t< t_i + 5.25}hr, and the acqua diamond spectrum 
signals the evolution during the decaying part. }
 \label{fig:FlareB_edist_spec}
\end{figure}

Encouraged by this combined MW spectroscopy/temporal success for Flare C, 
we moved to the more challenging case of the powerful GeV-band Flare B,
with results illustrated in Fig.~\ref{fig:FlareB_edist_spec}.  At its peak, 
the $\gamma$-ray flux of this flare is 3-10 times higher than that for Flare C.  While a simple 
increase in the shock acceleration injection rate \teq{Q_e(\gamma, \, t)} could 
yield this strong signal, it would also produce a similar flare enhancement in 
the optical synchrotron, which was clearly not seen.  Thus, to generate the 
pseudo-orphan GeV flare with a strongly Compton-dominated model, we lowered 
the jet frame magnetic field to \teq{B=0.075}Gauss,
(corresponding to a mildly-relativistic gyro-scale of  \teq{m_ec^2/eB = 2.27 \times 10^4}cm),
i.e. a factor of 8 or so below the \teq{B=0.65}Gauss employed for Flare C and
the long-term equilibrium MW spectra (dark blue curves in both 
Figs.~\ref{fig:FlareC_edist_spec} and~\ref{fig:FlareB_edist_spec}).  
This field reduction was ephemeral, with an exponential recovery on a timescale 
of just over 3 days, and can be interpreted as being part of larger scale MHD 
turbulence in the jet.  Such magnetic rarefactions followed by some recovery are seen 
in active regions just downstream of non-relativistic interplanetary shocks in the solar wind \cite{BOEF97}, 
albeit by factors of around \teq{1.5-2} in field strength.

The prominence of the EIC component in Flare B yielded a net flattening of the spectrum 
in the X-ray band as it dominates the SSC component.  This evolutionary spectral trend 
was commensurate with that of the {\it Swift/NuSTAR} flare data.  Another spectral characteristic 
was the blue-ward shift of the gamma-ray peak as the flare progresses, leading 
to distinctive spectral hysteresis properties that are depicted in \cite{BB19_ApJ} via 
familiar hardness-intensity diagrams.  To generate this trend, 
the model required an alteration of the turbulence/diffusion parameters from the quiescent, 
equilibrium values, and so we employed $\eta_1 = 100 \to 10$ and $\alpha = 3 \to 2.3$.
This modification is indicative of modest increases in turbulence levels in the post-shock
region \cite{SB12}.  The consequence was
to raise the value of \teq{\gammax} by a few.  This larger maximum Lorentz factor is 
apparent in the left panel of Fig.~\ref{fig:FlareB_edist_spec} during the injection phase, 
as is the rapid cooling in the relaxation phase that spawns hard-to-soft spectral evolution 
in both the optical and $\gamma$-ray bands.  Light curves reflecting this character for 
the frequencies marked on the SED panel are exhibited in Fig.~7 of \cite{BB19_ApJ}.
As with Flare C, the Flare B light curves in the X-ray and {\sl mm} radio peak after the cessation of 
acceleration/injection and decay much slower than do the flux time profiles for 
the optical synchrotron and EIC gamma-ray bands.  Our full paper \cite{BB19_ApJ}
graphically illustrates and discusses such inter-band lags for both flares.

No significant TeV emission is predicted by our modeling for either flare, in accordance
with the finding by \cite{Boettcher09} that leptonic models have difficulties reproducing the VHE emission 
observed in several exceptional flare states of 3C~279.  The cause of this 
is the low value of \teq{\gammax}, imposed by the very strong Compton cooling in the emission region,
and required to generate the low synchrotron peak frequency.
No VHE gamma-ray emission was detected from 3C~279 during the 
2013--2014 flaring episodes discussed in \cite{Hayashida15}.
Yet there have been TeV-band detections by Atmospheric \v{C}erenkov Telescopes for this blazar, 
for example by MAGIC in 2006 \cite{Albert-2008-Sci}, data for which is included as archival VHE flux points (gray) 
in both Figs.~\ref{fig:FlareC_edist_spec} and~\ref{fig:FlareB_edist_spec}.  Modeling 
of flares evincing emission above 100 GeV may demand more modest injection increases 
during flares concomitant with reduced Compton cooling, such as was explored for the HBL Mrk 501 in 
\cite{BBS17,BB19_ApJ}.

The treatment of full thermal + non-thermal electron distributions in the shock 
acceleration modeling from \cite{SB12} employed in this investigation permits 
measures of the thermal electron number density \teq{n_e} in the jet plasma.  
This is not afforded by commonplace MW modeling that restricts considerations to purely 
non-thermal fiducial electron distribution functions.  Accordingly we evaluate 
the plasma frequency \teq{\wp} in our modeling, finding values of \teq{\wp = 8.4}MHz
for Flare C and  \teq{\wp = 8.3}MHz for Flare B.  These values, which for the range of models 
attainable in our analysis possess a precision to within a factor of a few, establish the approximate 
energization rates for acceleration processes that connect to inertial effects such 
as relativistic magnetic reconnection or Weibel-instability instigated
turbulent shock acceleration.  The B-field values ascertained from the 
MW spectral fits yielded cyclotron frequencies of \teq{\ecyc = 11.4}MHz
for Flare C and  \teq{\ecyc = 1.3}MHz for Flare B.  These measures define the 
rates \teq{d\gamma /dt \sim \ecyc/\eta (p)} of cyclotronic acceleration processes such as DSA
at relativistic shocks;
comparison with the plasma frequencies indicates that reconnection and \underline{efficient}
diffusive acceleration (i.e., for \teq{\eta \sim 1}) are more or less on a par for the plasma jet of 3C 279.
Yet, our MW spectroscopy demands inefficient SDA shock acceleration conditions, 
and similar inefficiency would be imposed upon magnetic reconnection scenarios.

\section{Conclusion}

In this short cameo offering, we present some core results from our
numerical analysis coupling Monte-Carlo simulations of diffusive shock 
acceleration with time-dependent radiation transfer in an internal-shock, 
leptonic scenario for flares from the FSRQ blazar 3C 279.  Our two-zone 
picture, with a small acceleration zone, in which both diffusive shock and shock drift acceleration
are active, and a larger
radiation/electron cooling zone, extends our prior work in \cite{BBS17} that laid the foundation 
for this unique combination approach.  Our exploration of distinctive 3C 279 flares 
here during its 2013-2014 active period generates MW spectroscopy and 
temporal evolution broadly consistent with the observations.  Moreover, it 
attains a consistency with the general conclusions we obtained in \cite{BBS17}
for blazars Mrk 501, BL Lac and AO 0235+164, in its ability to constrain 
core physics parameters of the jet, namely the cyclotron and plasma frequencies.
More complete details are laid out in \cite{BB19_ApJ}, setting the scene for the 
next stage in our program which will include probes of more complex temporal 
injection profiles, and application to other bright, flaring blazars.

\vspace{3pt}
\acknowledgements{We thank NASA for support for early parts of this research 
program through the Astrophysics Theory Program, grant NNX10AC79G.
MGB is also grateful for support from the NASA {\it Fermi} 
Guest Investigator Program through grant 80NSSC18K1711.
The work of M. B\"ottcher is supported by the South African 
Research Chairs Initiative (grant no. 64789) of the Department of Science and 
Innovation and the National Research Foundation\footnote{Any opinion, finding 
and conclusion or recommendation expressed in this material is that of the authors 
and the NRF does not accept any liability in this regard.} of South Africa.}

\end{document}